\documentclass[aps,prd, onecolumn, superscriptaddress, nofootinbib, notitlepage, floatfix, preprintnumbers,longbibliography]{revtex4-1}
\usepackage{epsfig}
\usepackage{latexsym,amsmath,amssymb,amstext}
\usepackage{float}
\usepackage{graphicx,xcolor}
\usepackage[paperwidth=210mm,paperheight=297mm,centering,hmargin=2cm,vmargin=2.5cm]{geometry}

\newcommand{\be}{\begin{equation}}
\newcommand{\ee}{\end{equation}}
\newcommand{\bea}{\begin{eqnarray}}
\newcommand{\eea}{\end{eqnarray}}

\newcommand{\tr}{{\rm tr\ }}

\newcommand\bef{\begin{figure}}
\newcommand\eef[1]{\label{fg:#1}\end{figure}}
\newcommand\beq{\begin{equation}}
\newcommand\eeq[1]{\label{#1}\end{equation}}
\newcommand\beqa{\begin{eqnarray}}
\newcommand\eeqa[1]{\label{#1}\end{eqnarray}}
\newcommand\bet{\begin{table}}
\newcommand\eet[1]{\label{tb:#1}\end{table}}

\newcommand\fgn[1]{Figure \ref{fg:#1}}
\newcommand\eqn[1]{Eq.\ (\ref{#1})}
\newcommand\scn[1]{Section \ref{sec:#1}}

\newcommand\tbn[1]{Table \ref{tb:#1}}

\begin{document}

\date{\today}

\title{Perturbative computation in a QED$_3$-inspired conformal abelian gauge model on the lattice
}

\author{Nikhil\ \surname{Karthik}}
\email{nkarthik.work@gmail.com}
\affiliation{Department of Physics, College of William \& Mary, Williamsburg, VA 23185, USA}
\affiliation{Thomas Jefferson National Accelerator Facility, Newport News, VA 23606, USA}
\author{Matthew\ \surname{Klein}}
\email{mklei036@fiu.edu}
\affiliation{Department of Physics, Florida International University, Miami, FL 33199}
\author{Rajamani\ \surname{Narayanan}}
\email{rajamani.narayanan@fiu.edu}
\affiliation{Department of Physics, Florida International University, Miami, FL 33199}

\begin{abstract}
We perform perturbative computations in a lattice gauge theory
with a conformal measure that is quadratic in a non-compact abelian gauge field and is nonlocal, as 
inspired by the induced gauge action in massless QED$_3$.
In a previous work, we showed that coupling fermion sources to the gauge model led to nontrivial conformal data in the correlation functions of fermion bilinears that are functions of 
charge $q$ of the fermion.
In this paper, we compute such gauge invariant fermionic observables to order $q^2$ in lattice perturbation theory with the same conformal measure. We reproduce the expectations for 
scalar anomalous dimension from previous estimates in dimensional 
regularization.
We address the issue of the lattice regulator dependence of the amplitudes of correlation functions.

\end{abstract}

\maketitle

\section{Introduction}

Numerical evidence~\cite{Karthik:2015sgq,Karthik:2016ppr,Karthik:2017hol} from recent works point to the 
scale-invariance of the parity invariant noncompact QED in three dimensions with $2N_f$ flavors of massless two-component fermions. 
Motivated by these numerical results and pioneering studies in perturbative QED that shows the presence of an infra-red fixed point~\cite{Appelquist:1985vf,Appelquist:1986qw,Appelquist:1986fd,Appelquist:1988sr} in the large-$N_f$ limit, a lattice gauge model was studied in~\cite{Karthik:2020shl} which was expected and numerically shown to be conformal at length scales much larger in units of lattice spacing. 
The gauge measure on an infinite lattice is given by
\be
[dA] e^{-S};\qquad S = \frac{1}{2} \sum_x \sum_{j,k=1}^3 F_{jk}(x)  \left[ \frac{1}{\sqrt{\Box}} F_{jk}\right](x) ;\qquad F_{jk}(x) = (\partial_j A_k)(x) - (\partial_k A_j)(x);\qquad \Box = \partial^\dagger_k \partial_k,
\ee
where $\partial_k$ is the lattice forward derivative. 
The lattice action is apparently nonlocal, but the rationale 
behind studying such an action was the possibility to mimic the 
most dominant piece of the gauge-action that is induced by 
the massless fermion determinant in QED$_3$.
The non-compact gauge field, $A_j(x) \in {\mathbb R}$, is on the link connecting $x$ and $x+\hat j$. 
To make the theory to be a $U(1)$ gauge theory, 
only observables constructed out of the $U(1)$ valued gauge links
given by 
\be
U_j(x) = e^{iq A_j(x)},
\ee
were measured. In the above equation, $q$ is an arbitrary real-valued charge. At ${\cal O}(q^2)$, the charge 
can be identified with $16/N_f$ in 
the large-$N_f$ limit of QED$_3$, and such an identification breaks down at higher orders of $q$ but the lattice model is well-defined nevertheless. Using such gauge links, one can 
define the so-called pure gauge observables such a Wilson loops and their correlators. For example, 
the expression for a planar rectangular Wilson loop of size $\ell\times t$; $\ell,t \in {\mathbb I}$, is
\be
q^2{\cal W}(\ell,t)=-\ln\left\langle \exp\left( iq \sum_{x\in \ell\times t} F_{ij}(x)\right)\right\rangle = \frac{q^2}{2\pi^3} \int_{-\pi}^\pi d^3 p \frac { \sin^2 \frac{p_1\ell}{2} \sin^2 \frac{p_2 t}{2}}{\sqrt{ \sum_{k=1}^3 \sin^2 \frac{p_k}{2}}} \left[ \frac{ 1}{\sin^2 \frac{p_1}{2}} + \frac{ 1}{\sin^2 \frac{p_2}{2}}\right].
\ee
The asymptotic conformal behavior (that only depends linearly on the aspect ratio of the Wilson loop)  
after eliminating a perimeter term is given by
\be
{\cal W}(\ell,t) -{\cal W}\left(\frac{\ell+t}{2},\frac{\ell+t}{2}\right) \sim -0.0820 \left( \frac{\ell}{t} + \frac{t}{\ell}\right);\qquad \frac{\ell}{t} \to\infty
\quad {\rm or} \quad \frac{t}{\ell} \to\infty,
\ee
and the constant obtained by numerically evaluating the integral is universal.

In addition to the pure-gauge observables, the conformal
behavior of fermionic observables was found to have 
nontrivial dependencies on $q$.  In order to define such fermionic observables and $n$-point functions, the partition
function of the lattice gauge model coupled to 
massless fermion sources $\psi^\pm$ in a parity-invariant manner was given by,
\be
Z(\bar\psi^\pm,\psi^\pm) = \int [dA] e^{-S(A) + \bar\psi^+ {\cal G} \psi^+ + \bar\psi^- {\cal G}^\dagger\psi^-},
\ee
where ${\cal G}$ is the lattice massless fermion propagator coupled to charge-$q$ gauge links. From this, the flavor triplet scalar ($\Gamma=1$) and vector ($\Gamma=\sigma_k$) operators can be defined as differential operators acting on $Z$:
\be
O^{\pm}_\Gamma(x) \equiv \frac{\partial}{\partial \bar\psi^\pm(x)}\Gamma \frac{\partial}{\partial \psi^\mp(x)};\qquad O^{0}_\Gamma(x) \equiv \frac{1}{\sqrt{2}}\left(\frac{\partial}{\partial \bar\psi^+(x)}\Gamma \frac{\partial}{\partial \psi^+(x)} + \frac{\partial}{\partial \bar\psi^-(x)}\Gamma \frac{\partial}{\partial \psi^-(x)}\right).
\label{opdef1}
\ee
Given a lattice Dirac operator, one can compute correlations functions of
scalar and vector 
operators, 
\be
S(q;x)= \langle O_1^+(0) O_1^-(x)\rangle,\qquad  V_{ij}(q;x) = \langle O^+_{\sigma_i}(0) O^-_{\sigma_j}(x)\rangle
\ee
respectively, as examples of gauge invariant correlators. The separation $x$ will be integer valued and for $|x|\gg 1$ on an infinite lattice, the correlators will be given by
\be
S(q;x) \sim \frac{C_S(q)}{ |x|^{4-2\gamma_S(q)}};\qquad V_{ij}(q;x) \sim \frac{ C_V(q) \left(\delta^{ij} - 2\frac{x^i x^j}{x^2}\right)}{ |x|^4}.
\ee
Numerical analysis of the lattice conformal model~\cite{Karthik:2020shl} studied over a range of $q$ resulted in fits of the form
\be
\gamma_S(q) = 0.076(11) q^2 + 0.0117(15) q^4 + {\cal O}(q^6);\qquad \frac{C_V(q)}{C_V(0)}  = 1 - 0.0478(7) q^2 + 0.0011(2) q^4 + {\cal O}(q^6).\label{numres}
\ee
The coefficient of the leading term in $\gamma_S(q)$ from the 
lattice regularized method is consistent with $\frac{2}{3\pi^2}$ obtained in~\cite{Chester:2016ref} using continuum perturbation theory with a dimensional regularization based ultra-violet cutoff. On the other hand the coefficient of the leading correction to $C_V(q)$ is not consistent with a computation in continuum perturbation theory using dimensional regularization~\cite{Giombi:2016fct}, namely, $\frac{C_V^d(q)}{C_V^d(0)} = 1+\left( \frac{23}{9\pi^2} -\frac{1}{4}\right)q^2 + \cdots$. 
In addition to correlators, the $ L^{-1-\gamma_S}$ type finite size scaling of the 
low-lying eigenvalues $\Lambda_i$ of the Hermitian operator, $-i{\cal G}$,  on large enough $L^3$ boxes also give information on the scalar
scaling dimension $\gamma_S$.

This paper is a follow-up to the numerical work that we summarized above. The aim of this work is two-fold. Namely, (a) the 
observation that the nonperturbative lattice results for various 
quantities were empirically found to be power expandable as a series in $q$ that is rapidly convergent motivated us to develop a 
perturbative framework for the lattice regulated model to avoid 
Monte Carlo methods. This work develops the perturbative setup at ${\cal O}(q^2)$. The method presented can be developed further for higher-orders in $q$ and thereby with a possibility of performing interesting computations such as of the three-point function conformal data in the model at larger lattice sizes than practically possible in a Monte Carlo computation. (b) Unlike a typical lattice QFT with a well defined free-field-like UV continuum limit that removes any lattice regulator dependencies (and with a possible conformality at long-distances), the behavior of the present lattice model is different. As noted above, the conformality in the lattice regulated model automatically emerges in the long-distance limits of correlation functions and finite size scaling of eigenvalues. However, due to the absence of a UV continuum limit, it is not immediately clear which of the conformal data are universal with respect to the lattice regulator (e.g., type and parameters of lattice Dirac operator). In this work, within the perturbative framework, we address this question.

\section{Lattice perturbation theory}
The perturbation theory computation will be on a $L^3$ lattice. The gauge field will obey periodic boundary conditions and the gauge fixed action with a source term for the gauge fields  is
\be
S = \frac{1}{L^3 } \sum'_p \sum_{jk} \tilde A^*_j(p) \frac{ \Box^2(p) \delta_{jk} +\left(\frac{1}{\xi}-1\right)h_j(p)h^*_k(p)}{g^{2(1-n)}\Box^n(p)} A_k(p) 
+\sum_{x,k} J_k(x) A_k(x)
\label{gfactionp}
\ee
where the prime over the sum implies that $p=0$ is excluded; the Fourier transforms are defined by
\be
\tilde A_j(p) =  \sum_x   A_j(x) e^{i\frac{2\pi x\cdot p}{L} } ; \quad  \tilde A_j(p+L) = \tilde A_j(p);\quad \tilde A_j(0) =0;\quad A^*_j(p) = A_j(-p);
 \quad p_k \in [0,L-1];\quad k=1,2,3;
\ee
and 
\be
h_k(p) = e^{-i \frac{2\pi p_k}{L}}-1; \qquad  \Box(p) = 2\sqrt{\sum_k \sin^2 \frac{\pi p_k}{L}}.
\label{shkp}
\ee
The lattice model is conformal when $n=1$; the usual Maxwell action when $n=0$ and a gauge action for a Thirring model when $n=2$. The gauge fixing term maintains the conformal nature when $n=1$. The generating functional for computing gauge field correlations is
\bea
&& Z(J) = \exp \left[ \frac{1}{2} \sum_{x,y} \sum_{jk} J_j(x) G_{jk}(x-y) J_k(y)\right];\cr
&& G_{jk}(x) = \frac{1}{L^3} \sum_{p} \tilde G_{jk}(p)  e^{-i\frac{2\pi x\cdot p}{L} };\qquad
\tilde G_{jk}(p) = \frac{ \Box^2(p) \delta_{jk} -(1-\xi) h_j(p) h^*_k(p)}{2 g^{2(n-1)}\Box^{4-n}(p)}.
\eea
It is sufficient to perform the perturbation theory with overlap fermions~\cite{Karthik:2016ppr} to compare with \eqn{numres}. To this end, we provide the pertinent details for Wilson fermion kernel followed by details for overlap fermions in the next two sub-sections. 
 
\subsection{Wilson fermion kernel}\label{sec:wilson}

Fermions will obey anti-periodic boundary conditions and
the Wilson fermion operator, $D$, is defined as
\be
D(x_1,x_2) = 3\delta_{x_2,x_1}-\sum_i \left[ p_{i+} e^{iq A_i(x_1)} \delta_{x_2,x_1+\hat i} + p_{i-} e^{-iq A_i(x_2)} \delta_{x_2,x_1-\hat i} \right];
\qquad p_{i\pm}  = \frac{1\mp \sigma_i}{2}.
\ee
In order to perform perturbation theory, we write
\be
D(x_1,x_2) = D_0(x_1,x_2) + D_I(x_1,x_2)
\ee
where
\bea
D_0(x_1,x_2) &=& 3\delta_{x_2,x_1} -\sum_i \left[ p_{i+}  \delta_{x_2,x_1+\hat i} + p_{i-}  \delta_{x_2,x_1-\hat i} \right];\cr
D_I(x_1,x_2) &=& \sum_i \left[ p_{i+} t_{i+}(x_1) \delta_{x_2,x_1+\hat i} + p_{i-} t_{i-}(x_2)  \delta_{x_2,x_1-\hat i} \right];\qquad t_{i\pm}(x) =  \left[1-e^{\pm iq A_i(x)}\right].\label{d0di}
\eea
We will set up the perturbation theory computation in momentum space and
use the unitary transformation
\be
U(x,p) = \frac{1}{L^{\frac{3}{2}}} e^{-i\left[ \frac{2\pi x\cdot p}{L} + \frac{\pi x\cdot a}{L}\right]};\qquad a=(1,1,1)
\ee
to go between coordinate and momentum space.
The free fermion operator is
\be
 \tilde D_0(p_1,p_2) = \tilde D_0(p_1) \delta(p_1-p_2);\qquad \tilde D_0(p)  = 2\sum_k \sin^2 \left[\frac{\pi p_k}{L} +\frac{\pi}{2L}\right] - i \sum_k \sigma_k \sin \left[ \frac{2\pi p_k}{L} + \frac{\pi}{L}\right].
\ee
We write the interaction term as
\be
\tilde D_I(p_1,p_2) = - q\tilde D_1(p_1,p_2) -\frac{q^2}{2} \tilde D_2(p_1,p_2),
\ee
where
\bea
\tilde D_1(p_1,p_2) &=& \frac{i}{L^3} \sum_j W_{1j}(p_1,p_2) \tilde A_j^s(p_1-p_2);\qquad W_{1j}(p_1,p_2) = p_{j+} r_j(p_2) - p_{j-} r^*_j(p_1);\cr
\tilde D_2(p_1,p_2) &=& \frac{1}{L^3} \sum_j W_{2j}(p_1,p_2) \tilde A_j^c(p_1-p_2);\qquad W_{2j}(p_1,p_2) = p_{j+} r_j(p_2) + p_{j-} r^*_j(p_1).
\eea
and
\be
\tilde A^s_j(p) = \frac{1}{q} \sum_x \sin[qA_j(x)] e^{i\frac{2\pi x\cdot p}{L}};\qquad \tilde A^c_j(p) = \frac{2}{q^2} \sum_x \left(\cos[qA_j(x)]-1\right) e^{i\frac{2\pi x\cdot p}{L}};\qquad r_j(p) = e^{-i\left[ \frac{2\pi p_j}{L}+\frac{\pi}{L}\right]}.
\ee

\subsection{Overlap fermions}\label{sec:overlap}

Perturbation theory has been developed in the past for overlap fermions~\cite{Capitani:2002mp,Yamada:1998se}. Since it is not as well known as the one for Wilson fermions, we provide some technical details in the subsection.
The massless overlap Dirac operator is defined
by~\cite{Karthik:2016ppr}
\be
D_o = \frac{1+V}{2}\qquad V = X \frac{1}{\sqrt{X^\dagger X}};\qquad VV^\dagger=1;\qquad X = D-m_w;\qquad m_w \in (0,2).\label{overlapop}
\ee
The propagator is given by
\be
G_o = \frac{1-V}{1+V};\qquad G_o^\dagger = -G_o.\label{overlapprop}
\ee
We start by writing
\be
 X =  X_0 - q D_1 - \frac{q^2}{2} D_2 ;\qquad X_0 = D_0 - m_w;\qquad \frac{1}{\sqrt{X^\dagger X} }= Q_0 + q Q_1 + q^2 Q_2 + \cdots
\ee
and obtain
\bea
&& Q_0 = \frac{1}{\sqrt{X_0^\dagger X_0}};\qquad 
 Q_1\frac{1}{Q_0} + \frac{1}{Q_0} Q_1 =Q_0\left( X_0^\dagger D_1 + D_1^\dagger X_0\right) Q_0;\cr
&&  Q_2\frac{1}{Q_0} + \frac{1}{Q_0} Q_2 = - \frac{1}{Q_0} Q_1^2 \frac{1}{Q_0} + \left( Q_1\frac{1}{Q_0} + \frac{1}{Q_0} Q_1 \right)^2
+\frac{1}{2} Q_0  \left( X_0^\dagger D_2 + D^\dagger_2 X_0 -2 D_1^\dagger D_1\right) Q_0. \label{Qeqn}
\eea
If we write
\be
V= V_0 - 2qV_1 -2 q^2 V_2 + \cdots,
\ee
we can use \eqn{overlapop} and obtain
\be
V_0  = X_0 Q_0;\qquad V_1 = \frac{D_1 Q_0 - X_0 Q_1}{2};\qquad V_2 = \frac{D_2 Q_0 + 2D_1 Q_1 - 2X_0 Q_2}{4};\qquad \cdots.\label{Vpert}
\ee
The resulting perturbative expansion for the overlap propagator in \eqn{overlapprop} is
\be
G_o = G_e + q G_i V_1 G_i + q^2 G_i V_2 G_i + q^2 G_i V_1 G_i V_1 G_i + \cdots.
\ee
where
\be
G_e = \frac{1-V_0}{1+V_0};\qquad G_i = 1+G_e = \frac{2}{1+V_0};\qquad 
G_i^\dagger =  V_0 G_i = G_i V_0.
\ee

Upon going to momentum space,
\be
\tilde V_0(q_1,q_2) = \tilde V_0(q_1) \delta(q_1-q_2);\qquad \tilde V_0(q) = \frac{\tilde X_0(q)}{S_w(q)};\qquad \tilde X_0(q) = \beta(q) - i\sum_k \left( \sigma_k \sin \left[ \frac{2\pi q_k}{L} + \frac{\pi}{L}\right]\right)
\ee
where
\be
\beta (q) = 2\sum_k \sin^2 \left[\frac{\pi p_k}{L} +\frac{\pi}{2L}\right] - m_w;\qquad 
S_w^2(q) =  \beta^2(q) +  \sum_k \sin^2 \left[ \frac{2\pi q_k}{L} + \frac{\pi}{L}\right].
\ee
The external and internal free propagators are
\be
\tilde G_e(q) = \frac{ i\sum_k \left( \sigma_k \sin \left[ \frac{2\pi q_k}{L} + \frac{\pi}{L}\right]\right)}{S_w(q) + \beta(q)},
\qquad 
\tilde G_i(q) = \frac{ S_w(q) + \beta(q) + i\sum_k \left( \sigma_k \sin \left[ \frac{2\pi q_k}{L} + \frac{\pi}{L}\right]\right)}{S_w(q) + \beta(q)},
\ee
respectively.
The expression for $V_1$ in momentum space is given by
\bea
\tilde V_1(q_1,q_2) &=&  \frac{ i}{2L^3}\sum_j V_{1j}(q_1,q_2) \tilde A^s_j(q_1-q_2)\cr
V_{1j}(q_1,q_2) &=& \frac{W_{1j}(q_1,q_2) + \tilde V_0(q_1) W_{1j}^\dagger(q_2,q_1) \tilde V_0(q_2)}{\left[ S_w(q_1) +S_w(q_2)\right]}.
\eea
The expression for $V_2$ in momentum space is given by
\bea
\tilde V_2(q_1,q_2) &=& \frac{1}{2L^3} \left[ \sum_j \left\{  -V_{2j}(q_1,q_2) \tilde A^c_j(q_1-q_2)\right\}
+  \frac{1}{L^3} \sum_{q_3,j,k} \left\{ V_{2jk}(q_1,q_2,q_3)\tilde A^s_j(q_1-q_3) \tilde A^s_k(q_3-q_2) \right\}\right]\cr
V_{2j}(q_1,q_2)&=&\frac{- W_{2j}(q_1,q_2) + \tilde V_0(q_1)  W_{2j}^\dagger(q_2,q_1) \tilde V_0(q_2) }{2\left[ S_w(q_1)+S_w(q_2)\right]}\cr
V_{2jk}(q_1,q_2,q_3) &=& \frac{ \left[ \tilde X_0(q_1) W_{1j}^\dagger(q_3,q_1) - W_{1j}(q_1,q_3) \tilde X^\dagger_0(q_3) \right]  \tilde X_0(q_3) \left[ X_0^\dagger(q_3) W_{1k}(q_3,q_2) - W_{1k}^\dagger(q_2,q_3) X_0(q_2)\right]}{S_w^2(q_3) \left[ S_w(q_1)+S_w(q_2)\right]\left[S_w(q_1)+S_w(q_3) \right]\left[S_w(q_3)+S_w(q_2)\right]}\cr
&& + \frac{\tilde V_0(q_1) W^\dagger_{1j}(q_3,q_1) \tilde V_0(q_3) W^\dagger_{1k}(q_2,q_3) \tilde V_0(q_2)}{S_w(q_3)\left[ S_w(q_1)+S_w(q_2)\right]} 
\eea

\subsection{Gauge correlation functions}\label{sec:gaugeprop}

We will need to compute correlation functions that involve $\tilde A_j^s(p)$ and $\tilde A_j^c(p)$.
Noting that $\tilde A_j^s(p)$ is odd in the gauge field and $\tilde A_j^c(p)$ is even in the gauge field, even powers of $\tilde A_j^s(p)$ with any power of $\tilde A_j^c(p)$ will result in non-zero correlation functions. All of them will have a power series in $q^2$. For our purpose, we only need
\be
\langle \tilde A_{j}^c(p)\rangle = - L^3 G^c(0) \delta(p);\qquad 
G^c(0) = \frac{2L^3}{q^2} \left[  1- e^{-\frac{q^2}{2} g(0)} \right];\qquad
g(0) = \frac{2+\xi}{3L^3g^{2(n-1)}} \sum'_p \frac{1}{\Box^{2-n}(p)}, \ee
and
\be
\langle \tilde A_{j_1}^s(p_1) \tilde A_{j_2}^s(p_2) \rangle = L^3 \tilde G_{j_1j_2}^s(p_1) \delta(p_1+p_2);\qquad
\tilde G_{jk}^s(p) =
   \frac{1 }{q^2} e^{-q^2 g(0)}  \sum_{x} \sinh[q^2 G_{jk}(x)] 
e^{i\frac{2\pi x\cdot p}{L}}.
\ee
Note that
\be
\tilde G^s_{jk}(-p) = \tilde G^s_{kj}(p) = \left[G^s_{jk}(p)\right]^*.
\ee
The compactness of the gauge field coupled to fermions have been maintained in obtaining the above correlation functions. Since gauge invariance in perturbation theory is only valid order by order in $q^2$, the above correlation functions have be expanded in $q^2$ to extract gauge invariant coefficients.

\section{Meson correlation function}\label{sec:mescor}

The fermion operator discussed in \scn{overlap} acts on two component fermions. We will assume that we have two copies of two component fermions,
with the associated operators, $D_o$ and $D_o^\dagger$. We will be interested in meson correlation functions. With this mind let us associate two component fermions,
$\psi, \bar\psi$; with the operator $D_o$ and another set of two component fermions, $\chi, \bar\chi$; with the operator $D_o^\dagger$.
Let us denote the propagators by
\be
\langle \psi(x_1) \bar\psi(x_2) \rangle = G_o(x_1,x_2);\qquad
\langle \chi(x_1) \bar\chi(x_2) \rangle = - G_o(x_2,x_1)
\ee
and we have used \eqn{overlapprop}. 
Type of mesons we will consider are
\be
O_i(x) = \bar\psi(x) \Gamma_i \chi(x);\qquad \bar O_i(x) = \bar\chi(x)\Gamma_i \psi(x)
\ee
where $\Gamma_i = 1, \sigma_i$.
To be clear, as the theory does not have dynamical fermions per se, the 
above equation in terms of fermion operators is 
actually made rigorous in terms of fermion sources 
as discussed in \eqn{opdef1}.
The correlation functions are
\be
M^{ij}(x_1,x_2)=\langle \bar O_i(x_1) O_j(x_2) \rangle = \langle \bar\chi(x_1) \Gamma_i \psi(x_1) \bar\psi(x_2) \Gamma_j\chi(x_2) \rangle
 = \tr \left[ \Gamma_i G_o(x_1,x_2) \Gamma_j G_o(x_1,x_2)\right] 
 \ee
 where the trace is only on the spin indices. A transformation to momentum space yields
 \be
 \tilde M^{ij}(p_1,p_2)=  \frac{1}{L^3} \sum_{q_1,q_2} \tr \left[ \Gamma_i \tilde G_o(q_1,q_2) \Gamma_j \tilde G_o(q_1-p_1,q_2-p_2)\right].
\ee
Integrating over the gauge fields results in
\be
\tilde M^{ij}(p_1,p_2) = \tilde M^{ij}(p) \delta(p_1,p_2),
\ee
where
\be
\tilde M^{ij}(p) = \tilde M^{ij}_0(p)+ q^2\left[ \tilde M^{ij}_{1t}(p) + \tilde M^{ij}_{1d}(p) + \tilde M^{ij}_{1c}(p)\right] + O(q^4)\label{mescor}
\ee
where $\tilde M^{ij}_{1t}(p)$ is the tadpole term, $\tilde M^{ij}_{1d}(p)$ is the disconnected term and $\tilde M^{ij}_{1c}(p)$ is the connected term.
The leading term is
\be
\tilde M^{ij}_0(p) =   \frac{1}{L^3} \sum_q \tr \left [ \Gamma_i \tilde G_e(q) \Gamma_j \tilde G_e(q-p) \right]  
\ee
In order to compute the tadpole term we note that upon gauge averaging
\be
\langle V_2(q_1,q_2) \rangle 
=  O_2(q_1)  \delta(q_1-q_2);\qquad
O_2(q) = \frac{G^c(0)}{2}  \sum_j V_{2j}(q,q)  + \frac{1}{2L^3} \sum_{r,j,k} \left(V_{2jk}(q,q,r) \tilde G^s_{jk}(q-r)\right),
\ee
and this leads to
\be
\tilde M_{1t}(p) 
= \frac{1}{ L^3} 
\sum_{q}\tr \left[ \tilde G_i(q) O_2(q) \tilde G_i(q) \left( \Gamma_j \tilde G_e(q-p) \Gamma_i + \Gamma_i \tilde G_e(q+p) \Gamma_j \right) \right].\label{ovtad}
\ee

In order to compute the disconnected term, we note that upon gauge averaging
\bea
\sum_{q_3} \langle \tilde V_1(q_1,q_3) \tilde G_i(q_3) \tilde V_1(q_3,q_2) \rangle &=&
-\left[ \frac{1}{4L^3} \sum_{q_3,i_1,i_2} V_{1i_1}(q_1,q_3) \tilde G_i(q_3) V_{1i_2}(q_3,q_2) \tilde G^s_{i_1,i_2}(q_1-q_3)\right] \delta(q_1-q_2) \cr
&\equiv& - \tilde f_o(q_1)\delta(q_1-q_2),
\eea
and this leads to
\be
\tilde M_{1d}(p) = -\frac{1}{L^3} \sum_{q} \tr \left[ \tilde G_i(q) \tilde f_o(q) \tilde G_i(q) \left( \Gamma_j\tilde G_e(q-p) \Gamma_i +\Gamma_i \tilde G_e(q+p) \Gamma_j\right)\right].\label{ovdis}
\ee

The connected term is 
\bea
&&\tilde M_{1c}(p)\cr
&=& - \frac{1}{4L^6} \sum_{q_1,q_2,i_1,i_2} \tr\left[ \Gamma_i \tilde G_i(q_1) V_{1i_1}(q_1,q_2) \tilde G_i(q_2) \Gamma_j \tilde G_i(q_2-p) V_{1i_2}(q_2-p,q_1-p)
\tilde G_i(q_1-p) \right]G^s_{i_1,i_2}(q_1-q_2).
\cr&&\label{ovcon}
\eea

\subsection{Scaling of the numerical sums}

The dependence of the gauge propagators appearing in \scn{gaugeprop} appear in the exponents. Since gauge invariance is only assured to order $O(q^2)$ for the meson propagators, we expand $G^c(0)$ and $\tilde G^s_{j_1j_2}(p)$ to the leading order given by
\be
G^c(0) = L^3 g(0) +O(q^2);\qquad \tilde G^s_{jk}(p)  = \tilde G_{jk}(p).
\ee
We store the fermion and gauge propagators in momentum space for a fixed $L$ and this computation scales like $L^3$. 
Both the computation of $\tilde M^{ij}_0(p)$ for all $p$ and its Fourier transform to $M^{ij}_0(x)$ for all $x$ scale like $L^6$. 
The full computations of $\tilde O_2(q)$,  $\tilde f_o(p)$, $\tilde M^{ij}_{1t}(p)$, $\tilde M^{ij}_{1d}(p)$, $\tilde M^{ij}_{1t}(x)$ and $\tilde M^{ij}_{1d}(x)$ scale like $L^6$.

The computation of $\tilde M^{ij}_{1c}(p)$ for all $p$ scales like $L^9$ and this dominates the computational time. To reduce this computational time, we consider
two types of meson propagators in coordinate space, namely,
\be
 M^{ij}_z(x) = \frac{1}{L} \sum_{p} \tilde M^{ij}(0,0,p) e^{-i\frac{2\pi xp}{L}} \qquad {\rm and} \qquad  M_p^{ij}(x) = M^{ij}(0,0,x).\label{twocor}
 \ee
 These two correlators will be sufficient to study the asymptotic behavior of relevance. Since $\tilde M^{ij}_{1c}(0,0,p)$ will scale like $L^7$ our computation has been significantly reduced. Focussing on the expression for $\tilde M_{1c}(p)$  in \eqn{ovcon}, we note that
 \be
M_{1c}(x) = - \frac{1}{4L^6} \sum_{q_1,q_2,i_1,i_2} \tr\left[ \Gamma_i \tilde G_i(q_1) V_{1i_1}(q_1,q_2) \tilde G_i(q_2) \Gamma_j e^{-i\frac{2\pi x\cdot q_2}{L}} h_{i_2} (q_1-q_2,x)
\right]G_{i_1,i_2}(q_1-q_2),
\ee
where
\be
 h_{j} (q,x) =  \frac{1}{L^3} \sum_r  e^{i\frac{2\pi x\cdot r}{L}} \tilde G_i(r) V_{1j}(r,q+r)
\tilde G_i(q+r).
\ee
With the separation in coordinate space restricted to $(0,0,x)$, we note that both the computations of $h_{i_2}(q,x)$ and $\tilde M_{1c}(x)$ scales like $L^7$.

\section{Results from lattice perturbation theory}

Our aim is to extract the ${\cal O}(q^2)$ corrections to the anomalous dimensions and the two-point function amplitudes, which are $\gamma_S^1$, $C_S^1$ and $C_V^1$. 
To minimize computations, we will consider two correlators. In the first case, we will set the separation to an
on-lattice-axis value $x=(x_1,0,0)$, which we denote using 
a subscript $z$ as
\be
S_z(q;x_1) = \frac{C_s(q)}{|x_1|^{4-2\gamma_S(q)} };\qquad V_z(q;x_1) = \sum_{i=1}^3 V_{ii}(q;x_1) = \frac{C_V(q)}{|x_1|^4}.
\ee
Note that we have summed over all directions for the vector 
correlator above.
Assuming the scaling of correlators to be valid for all $x=(x_1,x_2,x_3)$,
we will also consider correlators at zero spatial momentum, denoted by a subscript $p$ as,
\be
S_p(q;x_1) = \int_{-\infty}^\infty dx_2 dx_3 S_z(q;x) = \frac{ \pi C_S(q)}{\left(1-\gamma_S(q) \right) |x_1|^{2-2\gamma_S(q)}};\qquad
V_p(q;x_1) = \int_{-\infty}^\infty dx_2 dx_3 V_z(q;x) = \frac{ \pi C_V(q)}{ |x_1|^{2}}.
\ee
Writing the anomalous dimension and the amplitudes order by order,
\be
\gamma_S(q) = \gamma^1_S q^2+\cdots;\quad C_S(q) = C^0_S + C^1_S q^2;\quad C_V(q) =  C^0_V +C^1_V q^2 + \cdots,
\ee
we have for ratios of correlators at non-zero $q$ with respect to
that in free field as
\bea
&& \frac{S_z(q;x_1)}{S_z(0;x_1)} =  1 + \left[ \frac{C_S^1 }{C_S^0}+ 2\gamma^1_S \ln|x_1|\right] q^2 \equiv 1 + q^2 R^z_S;\cr
&& \frac{S_p(q;x_1)}{S_p(0;x_1)} = 1 + \left[ \frac{C_S^1 }{C_S^0}+ \gamma_S^1+ 2\gamma^1_S \ln|x_1|\right] q^2 \equiv 1 + q^2 R^p_S;\cr
&&\frac{V_z(q;x_1)}{V_z(0;x_1)} = 1+ \frac{C_V^1 }{C_V^0} q^2 \equiv 1 + q^2 R^z_V; \cr
&&\frac{V_p(q;x_1)}{V_p(0;x_1)} = 1+ \frac{C_V^1 }{C_V^0} q^2 \equiv 1 + q^2 R^p_V.
\label{contratios}
\eea
with the equalities above valid only up to ${\cal O}(q^2)$. 
On a finite lattice of size $L^3$, all the quantities above 
have an implicit dependence on $L$ and one needs to 
perform $L\to\infty$ extrapolation at fixed $|x|$.  We will perform the following limits for the ratios
above as
\be
R^{z,p}_S(x_1) = \lim_{L\to\infty} R^{z,p}_S(x_1,L);\qquad 
R^{z,p}_V(x_1) = \lim_{L\to\infty} R^{z,p}_V(x_1,L).
\ee
using expansions in $x/L$ as
\be
 R^{z,p}(x,L) = R^{z,p N}_S(x) + \sum_{n=1}^N a^N_n(x) \left(\frac{x}{L}\right)^{2n}.\label{Lfit}
\ee
Since the fit is at a fixed $x$, grouping in powers of $x/L$ is just for convenience and a fit in even powers of $L$ is based on emperical observation.
We will use $N=7$ and $N=8$ to establish the stability of the leading term, $ R^{z,pN}(x)$. 
We computed the momentum sums on even lattices in the range $L\in [4,50]$. Keeping all $L > 2|x|$, we extracted the ratios at $L\to\infty$ for
$x_1\in [1,16]$. For sake of brevity, henceforth, we will denote 
the $x$-coordinate $x_1$ simply as $x$, and should not to be confused with 
the vector $x=(x_1,x_2,x_3)$ as in the discussion above.

\subsection{Scaling dimensions}

\begin {figure}
\includegraphics[scale=0.65]{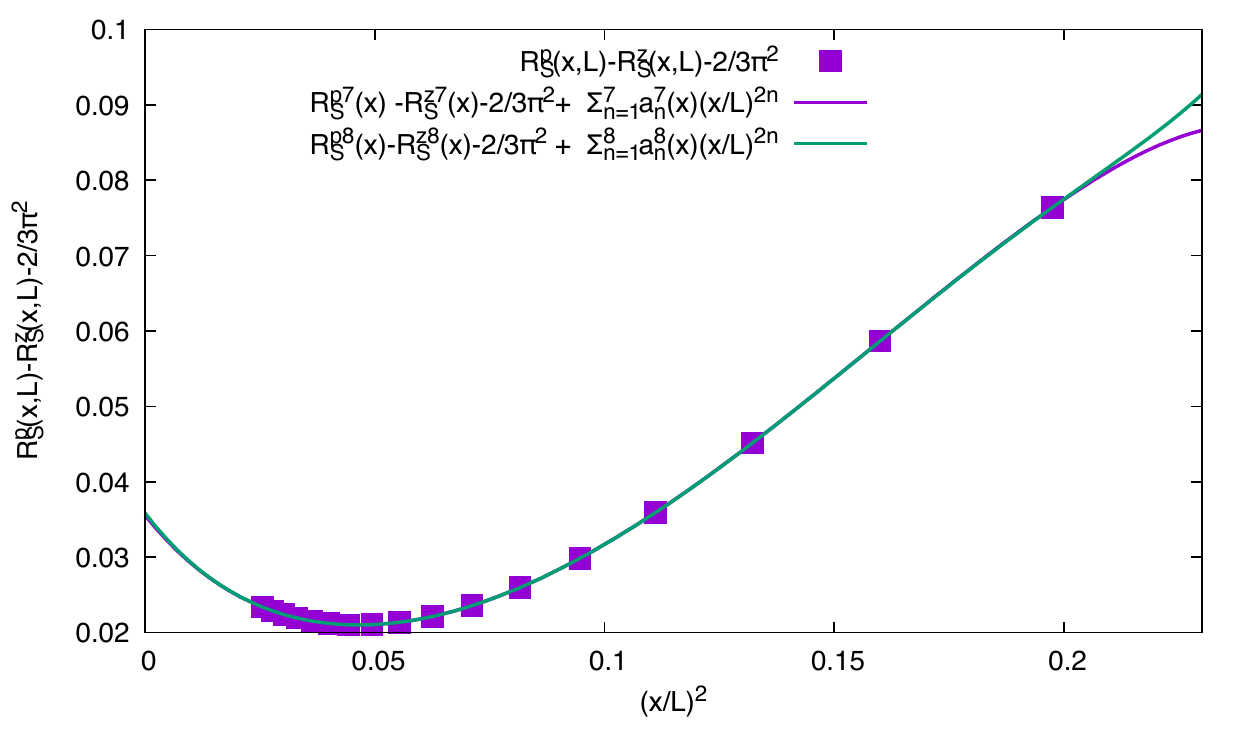}
\includegraphics[scale=0.65]{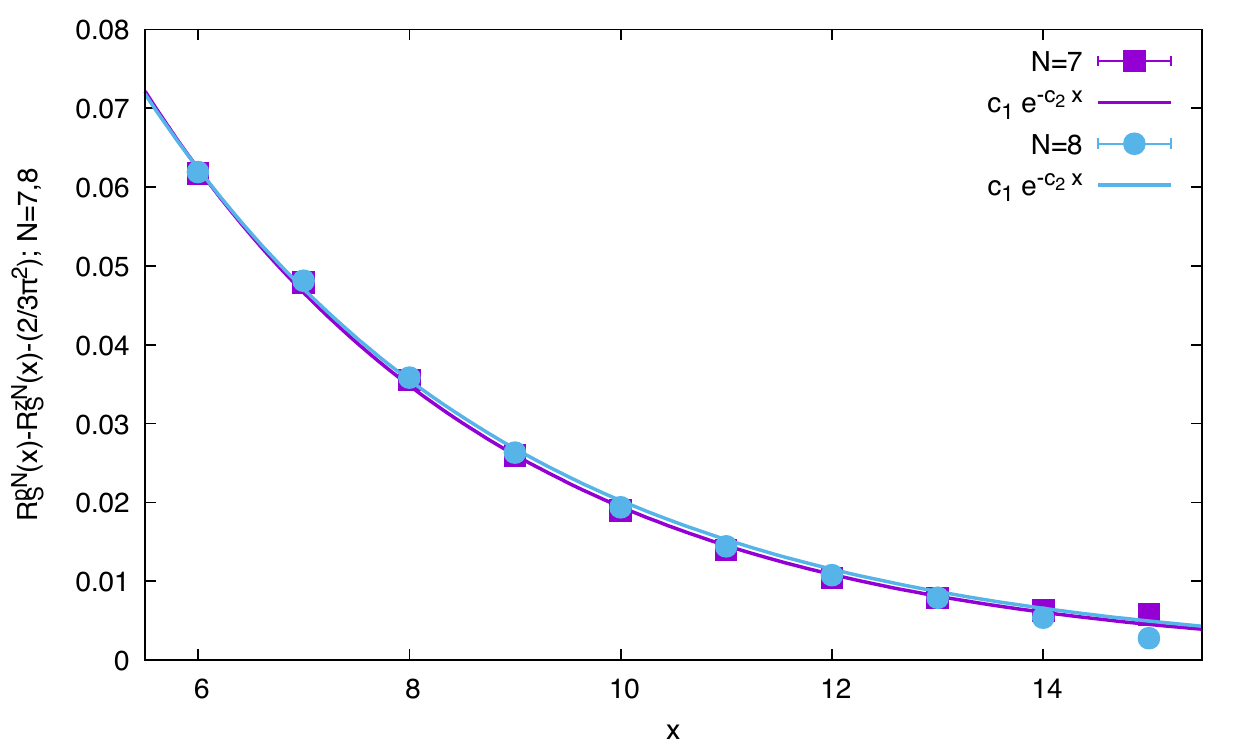}
\caption{Analysis details to study scaling dimension of the scalar using the difference between the zero spatial momentum correlator and 
the point-to-point correlator of the scalar meson using overlap fermion with $m_w=1.0$. The left  panel shows sample behavior of  $R_S^p(x,L)-R_S^z(x,L)-\frac{2}{3\pi^2}$  as a function of $\left(\frac{x}{L}\right)^2$ and the associated two different fits.  The value of $R_S^p(x)-R_S^z(x)-\frac{2}{3\pi^2} $ that is extracted for all values of $x\in[6,15]$ are shown along with the extrapolation errors in the right panel. The limit as $x\to\infty$, using single exponential fits of the type $c_1 e^{-c_2 x}$ shown as curves in the right panel,  
is consistent with zero.} \label{fg:ov-scalar-mom-coord}
\end {figure}

\begin {figure}
\includegraphics[scale=0.65]{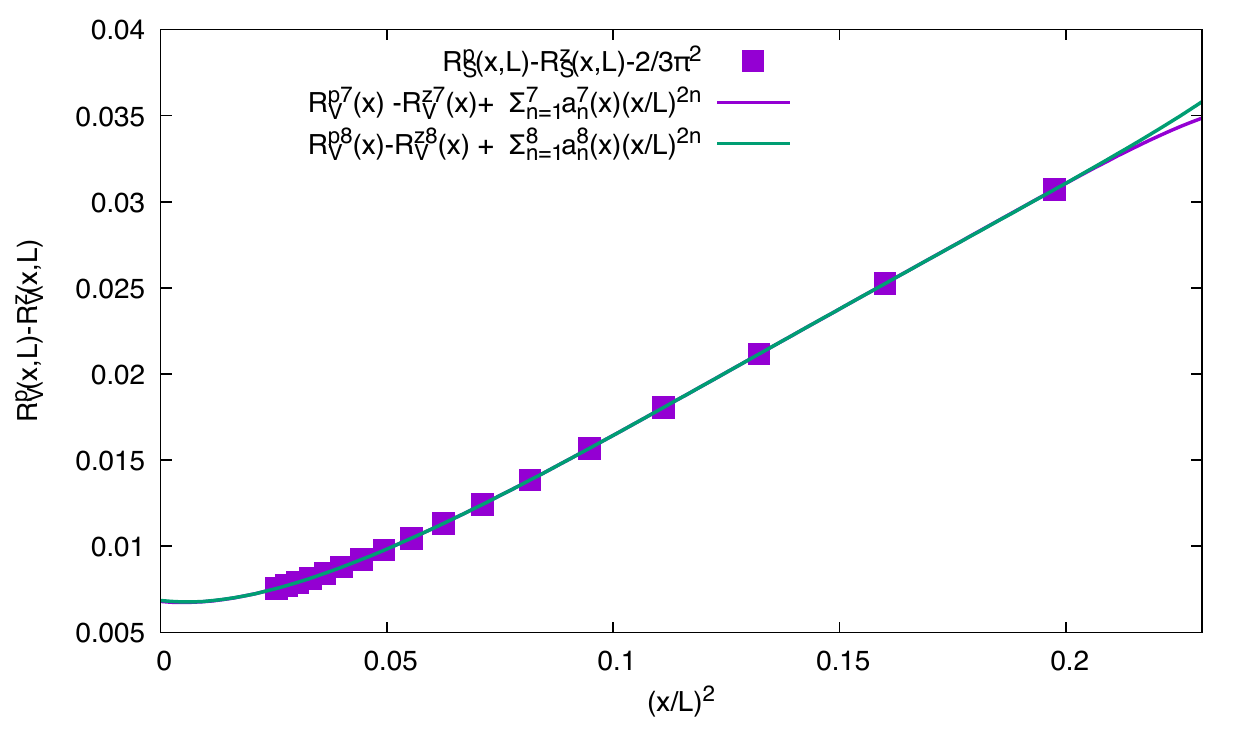}
\includegraphics[scale=0.65]{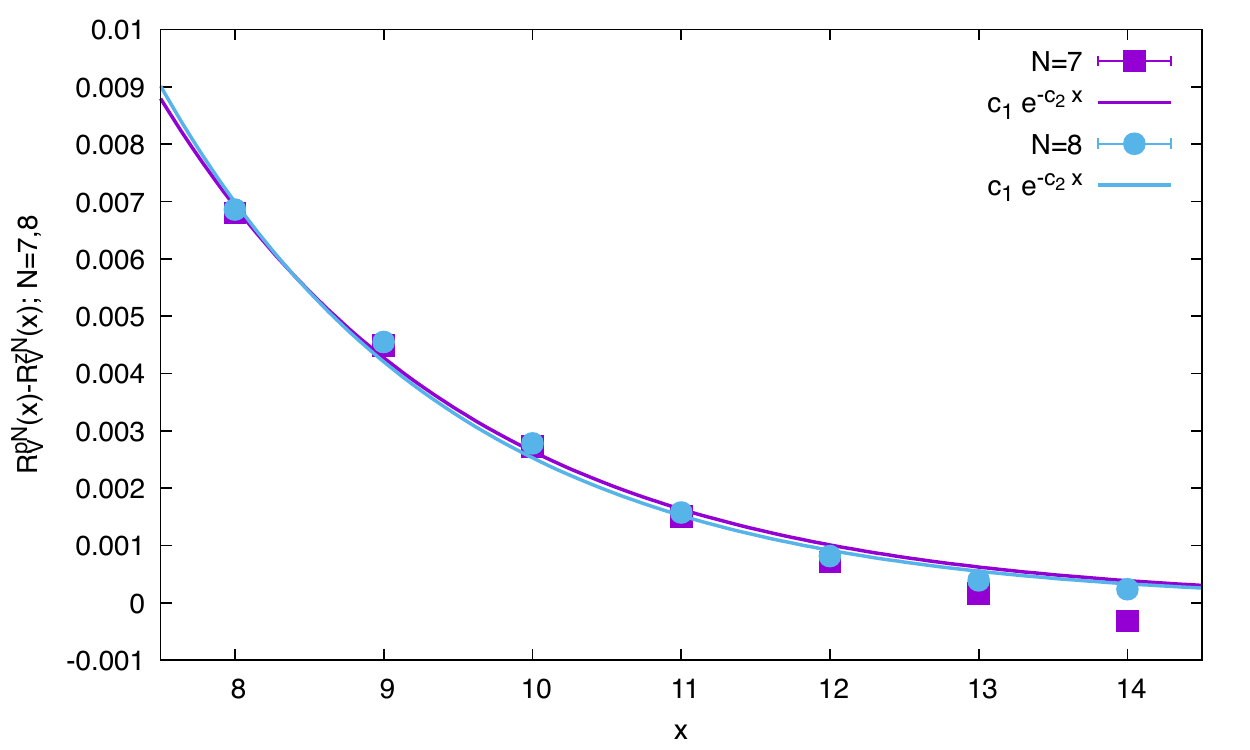}
\caption{Analysis details to study the absence of perturbative corrections to the scaling dimension of the vector using the difference of the zero spatial momentum correlator and the point-to-point correlator of the vector meson obtained using overlap fermion with $m_w=1.0$. The left  panel shows a sample behavior of  $R_V^p(x,L)-R_V^z(x,L)$  as a function of $\left(\frac{x}{L}\right)^2$ at $x=8$ and the associated fits with 
two different orders $N$.  The value of $R_S^p(x)-R_S^z(x)$ so extracted for all values of $x\in[8,14]$ are shown along with the extrapolation 
errors in the right panel. The limit as $x\to\infty$ using single exponential fit, as for the scalar case above, is consistent with zero.} \label{fg:ov-vector-mom-coord}
\end {figure}
We first concentrate on the scaling dimensions of
scalar and the vector within the lattice perturbation theory
at ${\cal O}(q^2)$, and can be obtained from 
$R_S$ and $R_V$ is the above equations. For the isotriplet vector,
one expects there to be no corrections from interaction to its 
free field scaling dimension.  The combinations,
\be
R_S^p(x) - R_S^z(x) \sim \gamma_S^1 q^2;\qquad R_V^p(x) - R_V^z(x) \sim 0 \label{smpcdiff},
\ee
for $|x|\gg 1$, can be seen to be good observables to extract the ${\cal O}(q^2)$ corrections to the scaling dimensions.

We study the above quantity for the scalar correlator using 
overlap fermion with $m_w=1.0$ in \fgn{ov-scalar-mom-coord}. 
From the dimensional regulatization computation, it is known that 
$\gamma_S^1=\frac{2}{3\pi^2}$. Therefore, we consider the 
combination $R_S^p(x,L) - R_S^z(x,L) - \frac{2}{3\pi^2}$.
The left panel shows
its behavior as a function of $\left(\frac{x}{L} \right)^2$ for 
a sample case of $x=8$. The infinite volume limits at each each fixed $x$ were obtained using 
the Ansatz of the type in \eqn{Lfit}. Such infinite volume extrapolated values at each $x$ with $N=7,8$ are plotted in the right panel as a function of $x$. It can be seen that the limit $x\to\infty$ is consistent with zero and a single exponential fit, $c_1 e^{-c_2 x}$,  matches the data reasonably well. Thus, we have shown that the result of 
$\gamma_S^1$ for the lattice model agrees with the expectation from dimensional regularization in the continuum at ${\cal O}(q^2)$.  
In addition to such a universality between continuum and lattice regulators, we also checked that the results for $\gamma_S^1$ from 
different $m_w$ in overlap fermion agree.

For the vector operator, we expect its scaling dimension to be 
uncorrected from the free field value to all orders in $q^2$. We demonstrate this using 
a similar strategy as for the scalar as shown in \fgn{ov-vector-mom-coord}. The left panel shows
the behavior of $R_V^p(x,L) - R_V^z(x,L) $ as a function of $\left(\frac{x}{L} \right)^2$ for $x=8$. The infinite volume extrapolated values at each $x$ with $N=7,8$ are plotted in the right panel as a function of $x$. Again, we find the limit $x\to\infty$ is consistent with zero and a single exponential fit matches the data reasonably well. The estimated value at $x=14$ from $N=7$ and $N=8$ fall on either side of zero. This implies that \eqn{smpcdiff} for the 
vector is reproduced without any regulator dependence.

\subsection{Two-point function amplitudes}

\subsubsection{Regulator dependence}

\begin {figure}
\includegraphics[scale=0.65]{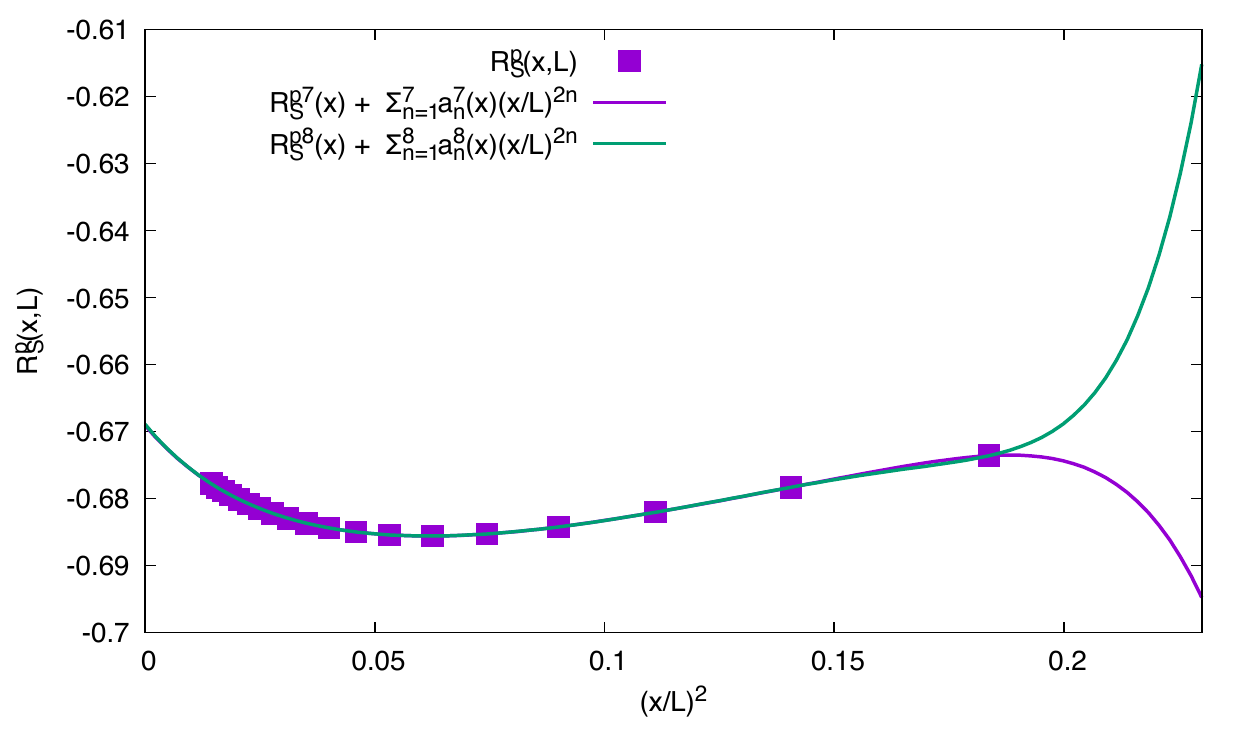}
\includegraphics[scale=0.65]{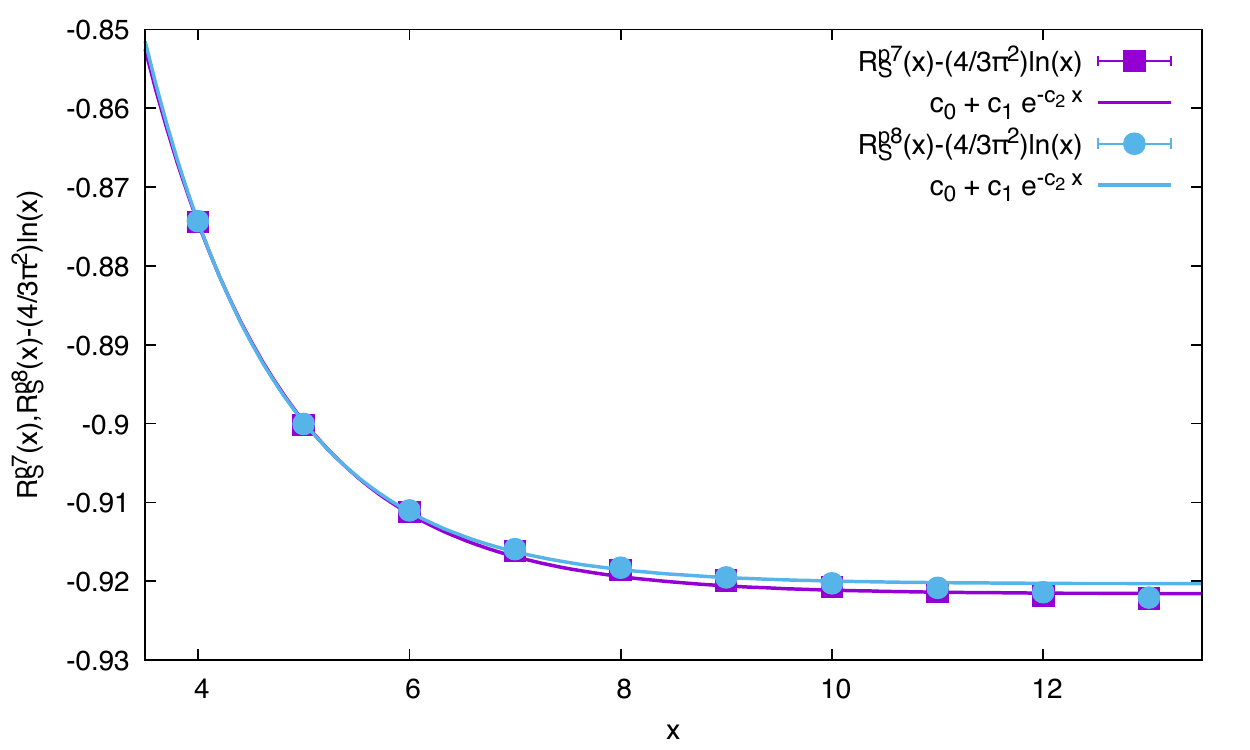}
\caption{Analysis details to obtain the scalar two-point function amplitude using overlap fermion with $m_w=0.5$. The left  panel shows sample behavior of $R_S^p(x,L)$ as a function of $\left(\frac{x}{L}\right)^2$ and the associated two different fits of the type in \eqn{Lfit} with $N=7$ and 8.  The value of  $R_S^p(x)-\frac{4}{3\pi^2} \ln(x) $ so extracted for all values of $x\in[4,13]$ are shown along with the extrapolation errors in the right panel. The single exponential fits to extract the amplitude in $x\to\infty$ limit are also shown as the curves.
} \label{fg:ov-scalar-mom}
\end {figure}

\begin {figure}
\includegraphics[scale=0.65]{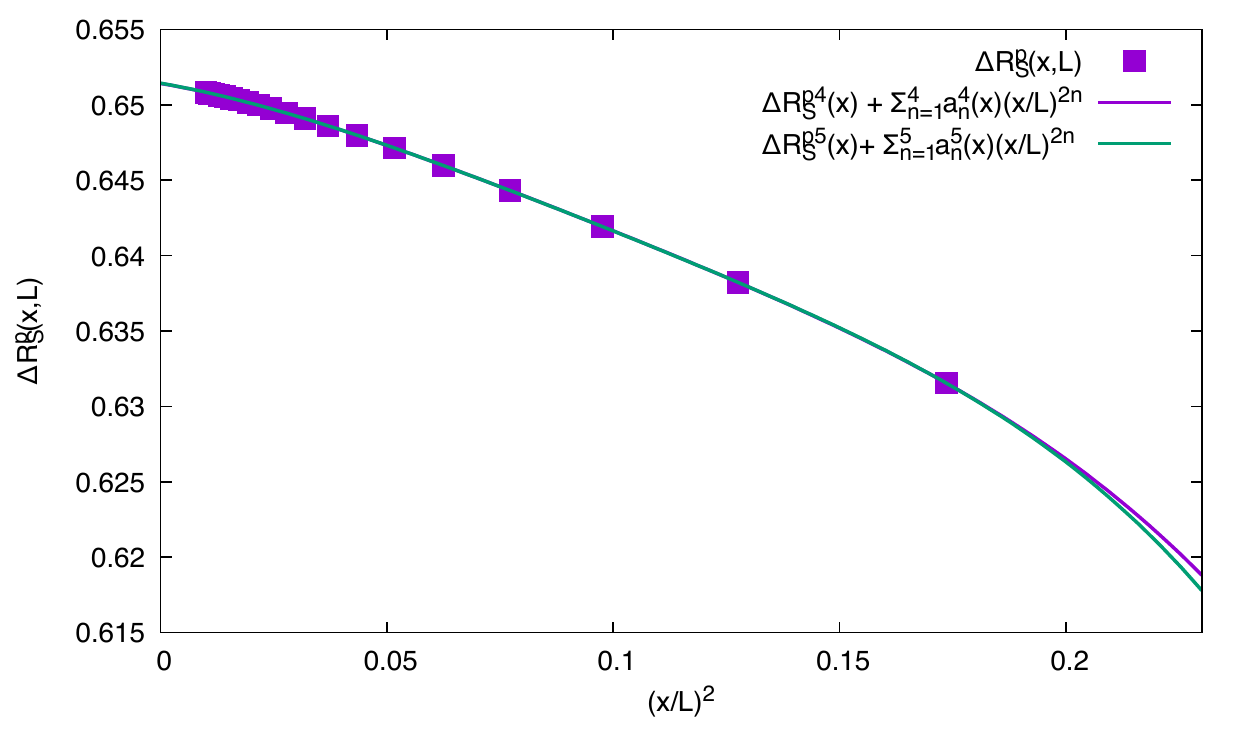}
\includegraphics[scale=0.65]{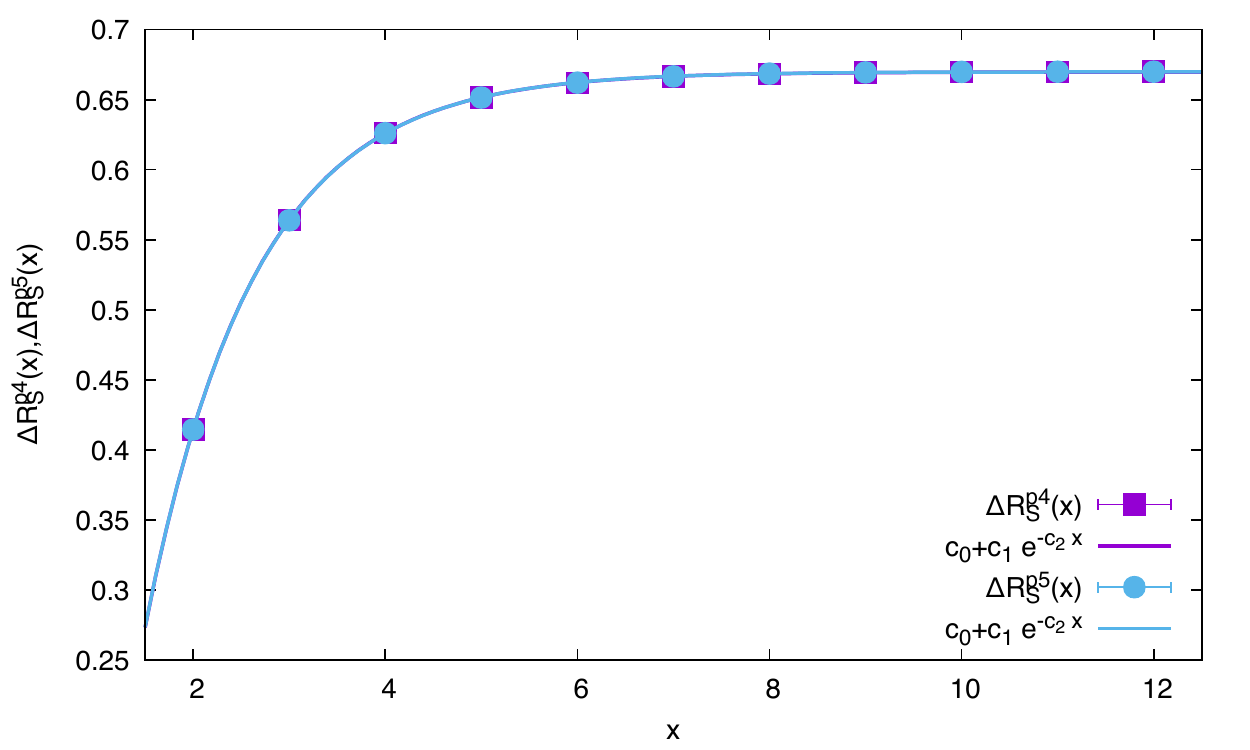}
\caption{A comparison of the results for overlap fermions with $m_w=0.5$ and $m_w=1.0$. The left panel shows a sample behavior of $\Delta R_S^p(x,L)$ as a function of $\left(\frac{x}{L}\right)^2$ and the associated two different fits.  The value of $\Delta R_S^p(x)$ so extracted for all values of $x\in[2,12]$ are shown along with the errors in the right panel.  
The limit as $x\to\infty$ is not zero and finite showing that the amplitude of the two-point function depends on the regulator parameter.
} \label{fg:diff-scalar-mom}
\end {figure}

We start our analysis by focussing on overlap fermion with $m_w=0.5$. The details are shown in \fgn{ov-scalar-mom}.
The left panel shows the data for $R_S^p(x,L)$  for overlap fermion with $m_w=0.5$. 
The data is plotted as a function of $\left(\frac{x}{L}\right)^2$ for a sample case of $x=6$. The extrapolated values at $L=\infty$ are $R_S^{p7}(6) = -0.66913$ and $ R_S^{p8}(6) = -0.66903$ and there is only a small systematic change in the fit values when one goes from $N=7$ to $N=8$.
Assuming that $\gamma_S^1 = \frac{2}{3\pi^2}$, we plot $R_S^p(x) - \frac{4}{3\pi^2} \ln x$ in the right panel for $N=7,8$ using
the infinite volume extrapolated values at different $x$. We see that the limit as $x\to\infty$ is finite and non-zero. A fit with a constant and single exponential fits the data well and we find that
\be
\frac{C_S^1}{C_S^0}\Bigg|_{ m_w=0.5} = -0.9885(6),
\ee
by comparing with \eqn{contratios}. The error in the numerical 
value on the right hand side of the above equation comes from 
the difference in the $N=7$ and $N=8$ values.

Next, we investigate the regulator dependence of the amplitude. To this end, we vary the Wilson mass parameter, $m_w$, within overlap fermions.
If the result is independent of the regulator, 
the difference in the results for  two different choices of $m_w$ should go to zero as $x\to\infty$. Let, 
\be
\Delta R_S^p(x,L)=R_S^p(x,L;m_w)-R_S^p(x,L;m_w=0.5),
\ee
denote the difference between two different regulators. Comparison of overlap fermion with $m_w=1.0$ to overlap fermion with $m_w=0.5$ is analyzed in \fgn{diff-scalar-mom}. 
The right panel shows the data for $\Delta R_S^p(x,L)$ where the difference is obtained by subtracting the ratio for overlap fermion with $m_w=0.5$ from overlap fermion with $m_w=1.0$. 
The data is plotted as a function of $\left(\frac{x}{L}\right)^2$ for $x=5$. A fit of the form in \eqn{Lfit}
with $N=4$ and $N=5$ are also shown. The extrapolated values at $L=\infty$ are $\Delta R_S^{p4}(5) = 0.651420$ and $\Delta R_S^{p5}(5) = 0.651440$, 
thereby showing only a small systematic dependence on the extrapolation ansatz.  
The systematic change in the fit values between the two choices of extrapolations is small.  The extrapolated values, $\Delta R_S^{p4}(x)$ and $\Delta R_S^{p5}(x)$, are plotted as a function of $x\in [2,12]$ in the right panel.  The $x\to\infty$ limit is approached exponentially and the data is fit using a constant and a single exponential. The limits are non-zero and finite, which clearly shows that the amplitude depends on the regulator parameter.
The dependence of the amplitude on $m_w$ are shown in the second column of \tbn{ov-scalar}. 

\begin{table}
\begin{tabular}{||c|c|c||}
\hline
$m_w$ & $\frac{C_S^1}{C_S^0}\Bigg|_{ m_w}  - \frac{C_S^1}{C_S^0}\Bigg|_{0.5}$ & Tadpole corrected result \\
\hline\hline
0.25 & -1.3328(37) & -0.1390(37) \\
\hline
0.75 & 0.44590(10) & 0.04797(10)\\
\hline
1.0 & 0.66976(6) & 0.07286(6)\\
\hline
1.25 &  0.80593(9) & 0.08965(9)\\
\hline
1.5 & 0.89843(43) & 0.10256(43)\\
\hline
1.75 & 0.9678(18) & 0.1151(18)\\
\hline
\end{tabular}
\caption{Table showing the dependence of the scalar meson amplitude ratio on the regulator for overlap fermions.
The second column is using the unimproved gauge links, and the third column is using 
tadpole improved gauge links (see text).} \label{tb:ov-scalar}
\end{table}


Our analysis of vector mesons mirrors the one for scalar mesons.
We start our analysis by focusing on overlap fermion with $m_w=0.5$ to extract the amplitude. The details are shown in \fgn{ov-vector-mom}.
The left panel shows the data for $R_V^p(x,L)$  for overlap fermion with $m_w=0.5$. 
The data is plotted as a function of $\left(\frac{x}{L}\right)^2$ for $x=6$. We needed to use 
$N=7$ and $N=8$ in \eqn{Lfit} (the form of fit is same for vector and scalar mesons) to best fit the data and these are also shown. The extrapolated values at $L=\infty$ are $R_S^{p7}(6) = -0.910487$ and $ R_S^{p7}(6) = -0.910450$.  
We plot $R_V^p(x)$ in the right panel for $N=7,8$. We see that the limit as $x\to\infty$ is finite and non-zero. A fit with a constant and single exponential fits the data well and we find that
\be
\frac{C_V^1}{C_V^0}\Bigg|_{ m_w=0.5} = -0.92254(13).
\ee
\begin {figure}
\includegraphics[scale=0.65]{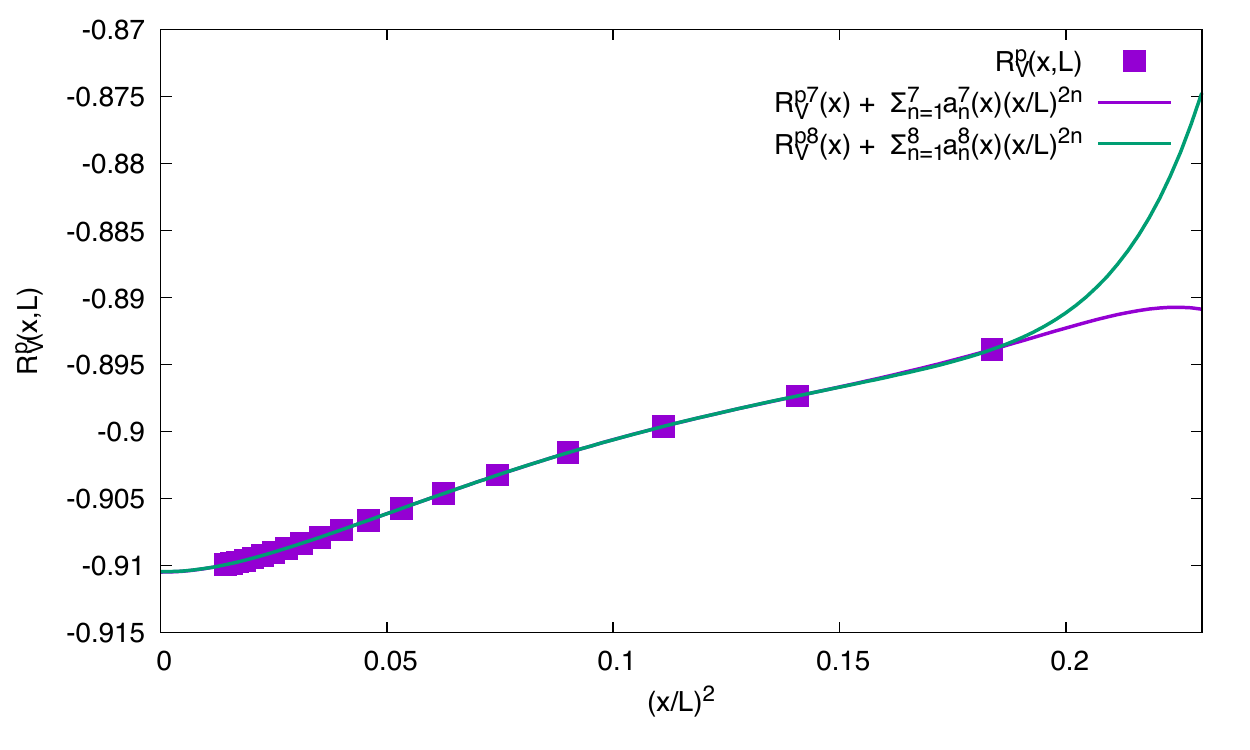}
\includegraphics[scale=0.65]{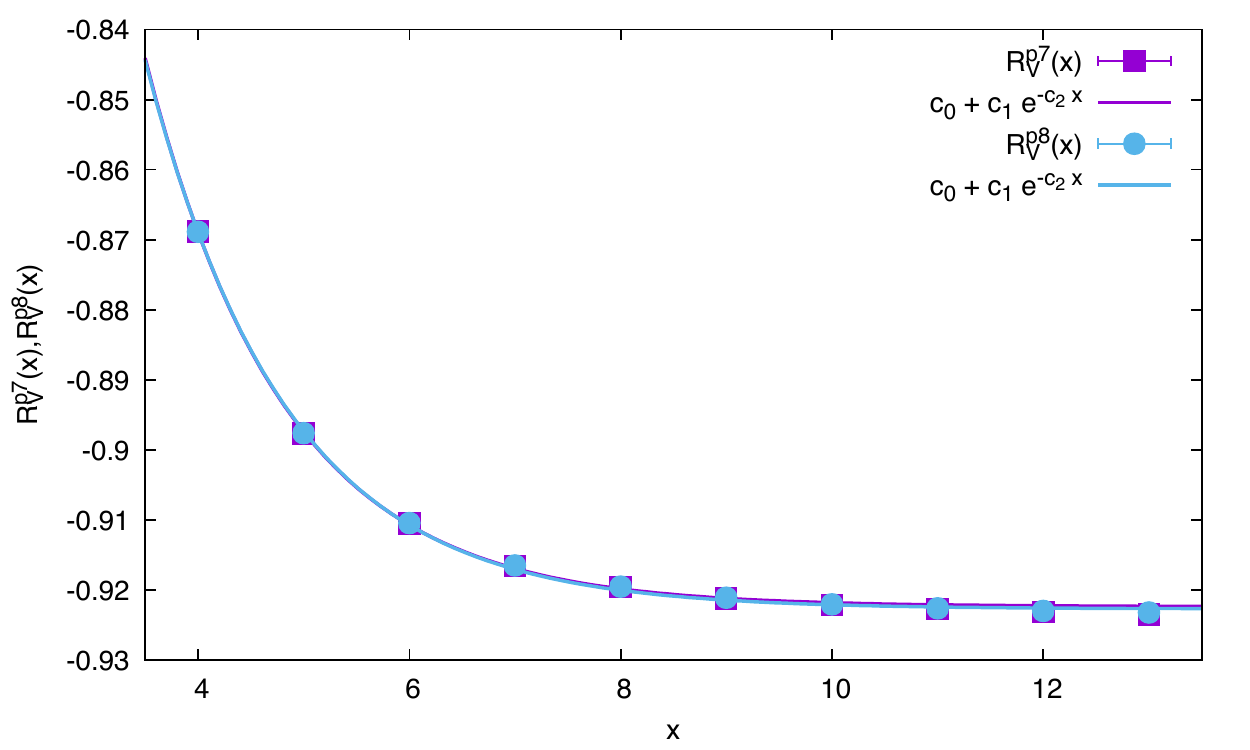}
\caption{Analysis details for the zero spatial momentum projected vector correlator using overlap fermion with $m_w=0.5$. The left  panel shows sample behavior of $R_V^p(x,L)$as a function of $\left(\frac{x}{L}\right)^2$ at a sample $x=5$, and the associated two different infinite volume extrapolation fits.  The value of  $R_V^p(x) $ so extracted for all values of $x\in[4,13]$ are shown along with the errors in the right panel. The fits to extract the leading correction to the amplitude are also shown.
} \label{fg:ov-vector-mom}
\end {figure}

\begin {figure}
\includegraphics[scale=0.65]{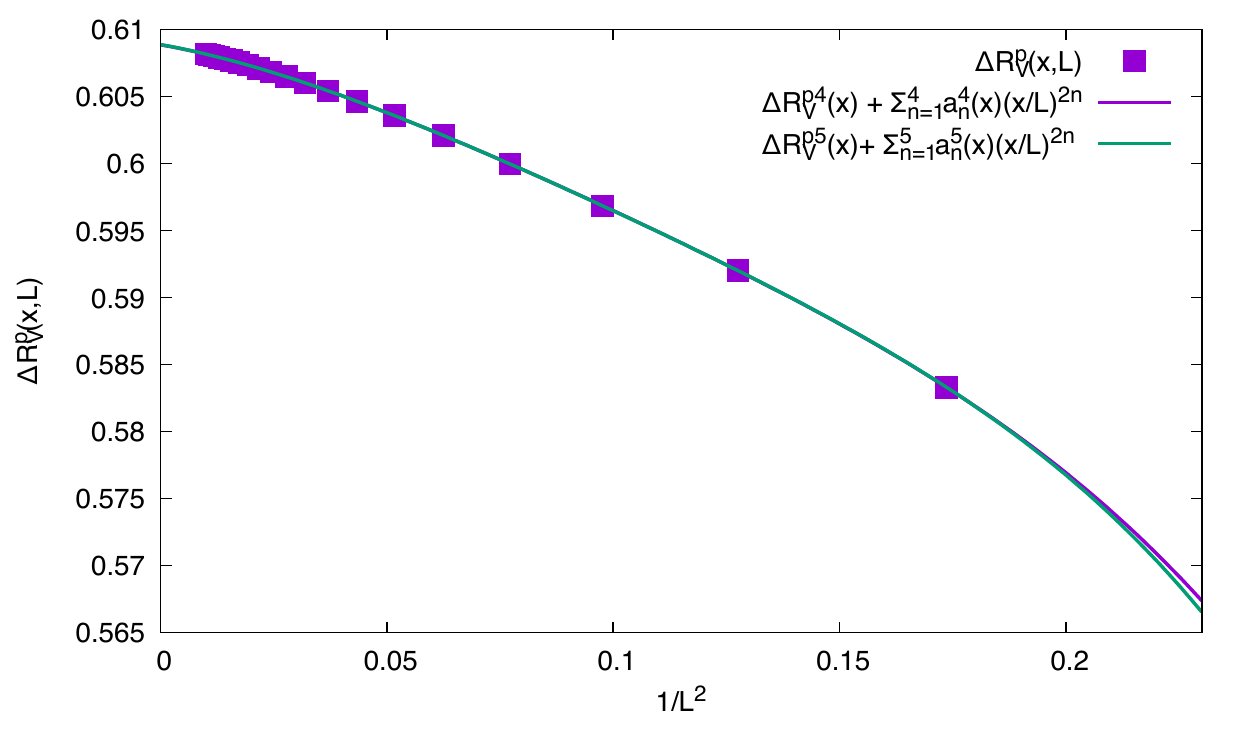}
\includegraphics[scale=0.65]{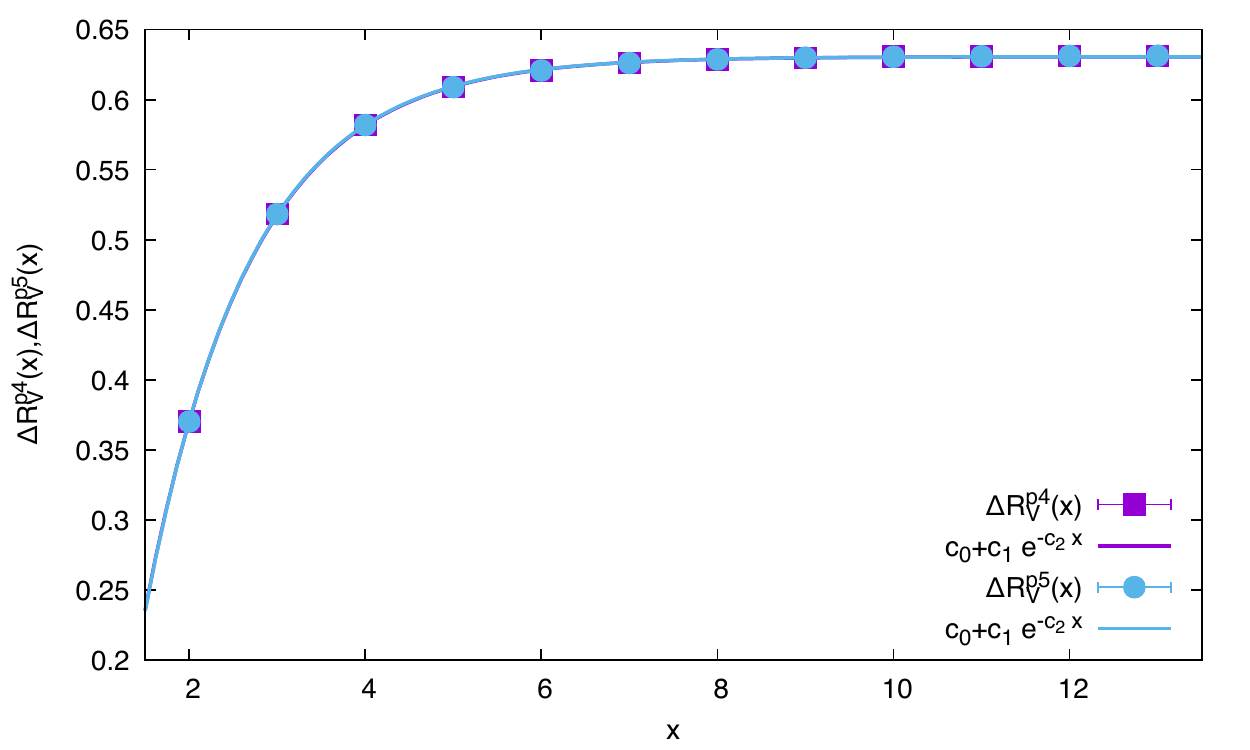}
\caption{
A comparison of the results for overlap fermions with $m_w=0.5$ and $m_w=1.0$. The left panel shows a sample behavior of $\Delta R_V^p(x,L)$ as a function of $\left(\frac{x}{L}\right)^2$ and the associated two different fits.  The value of $\Delta R_V^p(x)$ so extracted for all values of $x\in[2,12]$ are shown along with the errors in the right panel.  
The limit as $x\to\infty$ is not zero and finite showing that the amplitude of the two-point function depends on the regulator parameter.
} \label{fg:diff-vector-mom}
\end {figure}

Like in the case of scalar mesons, we investigate the regulator dependence of the amplitude by varying the Wilson mass parameter, $m_w$, within overlap fermions. Comparison of overlap fermion with $m_w=1.0$ to overlap fermion with $m_w=0.5$ is analyzed in \fgn{diff-vector-mom}. 
The right panel shows the data for $\Delta R_V^p(x,L)$ where the difference is obtained by subtracting the ratio for overlap fermion with $m_w=0.5$ from overlap fermion with $m_w=1.0$. 
The data is plotted as a function of $\left(\frac{x}{L}\right)^2$ for $x=5$. A fit of the form in \eqn{Lfit}
with $N=4$ and $N=5$ are also shown. The extrapolated values at $L=\infty$ are $\Delta R_V^{p4}(5) = 0.608857$ and $\Delta R_V^{p5}(5) = 0.608873$.   
We see only a small systematic change in the fit values when one goes from $N=4$ to $N=5$.  The extrapolated values, $\Delta R_V^{p4}(x)$ and $\Delta R_V^{p5}(x)$, are plotted as a function of $x\in [2,13]$ in the right panel.  The $x\to\infty$ limit is approached exponentially and the data is fit using a constant and a single exponential. The limits are non-zero and finite clearly showing that the amplitude of vector two-point function also depends on the regulator parameter.
The dependence of the amplitude on $m_w$ are shown in the second column of \tbn{ov-vector}. 

\subsubsection{Partial restoration of universality with tadpole improvement}
The regulator dependence of the two-point functions seen in \tbn{ov-scalar} and \tbn{ov-vector} in the 
lattice model is a curious aspect of this lattice gauge 
model, which approaches the continuum behavior simply at 
distance scales much larger than one lattice unit without any fine tuning. 
The regulator dependence of amplitudes is to be understood by the fact that the plaquette value 
in this model never approaches 1 due to the absence of the 
traditional continuum limit at a field field fixed point.
Thus, we wanted to check whether by ``improving" the 
Dirac operator by using gauge links that are closer to 
unity subdues the regulator dependence of the amplitudes.
A well known method to achieve this is via tadpole improvement,
namely, the replacement of 
the massless free Wilson-Dirac operator in \eqn{d0di} by
\be
D_0(x_1,x_2) = 3\delta_{x_2,x_1} -u_o\sum_i \left[ p_{i+}  \delta_{x_2,x_1+\hat i} + p_{i-}  \delta_{x_2,x_1-\hat i} \right]
\ee
where $u_0^4$ is the expectation value of the compact plaquette with charge $q$. A simple computation yields,
\be
u_0 = \exp \left[ -\frac{ q^2}{24 L^3} \sum'_p \Box(p)\right]
= e^{-\alpha q^2};\qquad \alpha=0.0994834.
\ee
This amounts to a change in the Wilson mass parameter by
\be
m_w \to \frac{ m_w -3(1-u_0)}{u_0}.\label{tadpolemw}
\ee
Since the free massless overlap propagator behaves
as
\be
\tilde G_e(q) = 2m_w \frac{ i\sigma_k p_k}{p^2};\qquad p_k = \frac{2\pi q_k}{L} \to 0
\ee
the induced wavefunction normalization is $\frac{1}{2m_w}$ for each fermion propagator. Since $m_w$ has a tadpole correction given by \eqn{tadpolemw}, we conclude that all ratios defined in \eqn{contratios} should be multiplied by
\be
 \left[ \frac{u_0}{1-\frac{3(1-u_0)}{m_w}}\right]^2  = \left[ 1 + \frac{2(3-m_w)\alpha}{m_w} q^2 +\cdots \right] .
\ee
This amounts to 
\be
\frac{C_{S,V}^1}{C_{S,V}^0} \to \frac{C_{S,V}^1}{C_{S,V}^0} + \frac{2(3-m_w)\alpha}{m_w}
\ee
resulting in 
\be
\frac{C_{S}^1}{C_{S}^0}\Bigg|_{ m_w=0.5}  + 10\alpha = 0.0063(6),
\ee
as the tadpole corrected amplitude ratio at $m_w=0.5$ 
and
\be
\frac{C_{S}^1}{C_{S}^0}\Bigg|_{ m_w} - \frac{C_{S}^1}{C_{S}^0}\Bigg|_{ m_w=0.5} + \frac{6(1-2m_w)\alpha}{m_w}\label{tadcorrsv}
\ee
as the tadpole corrected difference of the amplitude ratio. These are shown in the third column of \tbn{ov-scalar}.
Since the logic of the tadpole correction carries over to vector mesons, we can use \eqn{tadcorrsv} to include a tadpole correction resulting in
\be
\frac{C_{V}^1}{C_{V}^0}\Bigg|_{ m_w=0.5}  + 10\alpha = 0.07229(13),
\ee
and the third column in \tbn{ov-vector}.
In both the scalar and vector cases, the regulator dependence in the tadpole 
improved case is indeed weaker.

\begin{table}
\begin{tabular}{||c|c|c||}
\hline
$m_w$ & $\frac{C_V^1}{C_V^0}\Bigg|_{ m_w}  - \frac{C_V^1}{C_V^0}\Bigg|_{0.5}$ & Tadpole corrected result \\
\hline\hline
0.25 &  -1.3072(44) & -0.1134(44)\\
\hline
0.75 & 0.42461(7) & 0.02668(7)\\
\hline
1.0 & 0.630541(7) & 0.033641(6)\\
\hline
1.25 & 0.75052(7) & 0.03424(7)\\
\hline
1.5 & 0.8276(6) & 0.0317(6)\\
\hline
1.75 & 0.8824(18)& 0.0297(18)\\
\hline
\end{tabular}
\caption{Table showing the dependence of the vector meson amplitude ratio on the regulator for overlap fermions.
The second column is using the unimproved gauge links, and the third column is using 
tadpole improved gauge links (see text).
} \label{tb:ov-vector}
\end{table}

\section{Conclusions}

It is useful to compute corrections to conformal correlation functions  in a perturbation theory that maintains conformal invariance~\cite{Chester:2016ref,Giombi:2016fct}, with the possibility of performing $N$-point functions 
beyond $N=2$ on larger lattices without a Monte Carlo effort. Naively, only the anomalous scaling dimensions of operators and amplitudes of three point functions and higher (with the amplitudes of 2-point function set to unity as the normalization condition) are physical. There are situations that involve conserved operators where the amplitude of two point functions become physical. One such quantity is the vector current in conformal three dimensional QED. A lattice model to reproduce results in conformal three dimensional QED was proposed in~\cite{Karthik:2020shl}. We studied this model using lattice perturbation theory in this paper. We computed corrections to the scalar and vector two point functions. We showed that the scalar anomalous dimension is correctly reproduced and is independent of the regulator, thereby validating further future efforts within a lattice perturbation theory setup. On the other hand, we showed that the corrections to the amplitude of the scalar and vector two point function depends on the lattice regulator. In particular, we found that the amplitude of the vector correlator depends on the lattice regulator. 
This observation demands one to numerically revisit the verification~\cite{Karthik:2020shl} of the conjectured self-duality of three dimensional QED with four flavors of two component fermions~\cite{Wang:2017txt,Xu:2015lxa,Hsin:2016blu} within the framework of the lattice conformal model via the degeneracy of 
flavor current and topological current correlators; in the work~\cite{Karthik:2020shl}, the regulator dependence was not explored.
Since such a degeneracy between the correlators 
was also seen to arise within statistical errors in a conventional simulation of three dimensional QED~\cite{Karthik:2017hol} with a well-defined 
continuum limit,
we suspect that the value of $q$ in the lattice model 
where the flavor and vector currents coincide might turn out to be a universal value independent of the regulator. 
For this, one might need to use the induced Chern-Simons terms from massive fermions to compute the topological current correlator, wherein similar regulator dependence could be induced in the correlators of the fermion-based definition of the topological currents as well. Such a 
scenario conjectured by us needs to be studied further. In the future, it would also be interested to use the model to 
study scaling dimensions of monopoles by coupling the lattice model to the
gauge field $q A + {\cal A}_Q$, with $A$ being the 
dynamical gauge field and ${\cal A}_Q$
being the background gauge field for a flux $Q$ monopole-antimonopole pair as studied in~\cite{Karthik:2019mrr,Karthik:2018rcg}, and 
ask if they match the values found in different
$N_f$ flavor QED$_3$.
\acknowledgments
R.N. acknowledges partial support by the NSF under grant number
PHY-1913010. N.K. is supported by Jefferson Science
Associates, LLC under U.S. DOE Contract \#DE-AC05-
06OR23177 and in part by U.S. DOE grant \#DE-FG02-
04ER41302. 
\bibliography{biblio}
\end{document}